\documentclass[onecolumn,showpacs,preprintnumbers,amsmath,amssymb]{revtex4-1}

\usepackage{lipsum}
\usepackage{tabularx}
\usepackage{array}
\usepackage{mathrsfs}
\usepackage{amssymb}
\usepackage{amsmath}
\usepackage{cases}  % To number the subequations of an equation
\usepackage{graphicx}% Include figure files
\usepackage{subfigure} %To insert multiple figures together
\usepackage{dcolumn}% Align table columns on decimal point
\usepackage{multirow} % To merge multirow of the table
\usepackage{bm}% bold math
\usepackage{verbatim} % used for commenting on multiple paragraphs
\usepackage[dvipdfm,pdfstartview=FitH,colorlinks,linkcolor=blue,anchorcolor=blue,citecolor=blue]{hyperref}
\usepackage{amsthm}  %It is important for the proof environment

\begin{document}

\title{Dove prism in single-path Sagnac interferometer for orbital-angular-momentum photon states}

\author{Fang-Xiang Wang$^{1,2,3}$}
\author{Wei Chen$^{1,2,3}$}
\email{Corresponding author: weich@ustc.edu.cn}
\author{Ya-Ping Li$^{1,2,3}$}
\author{Guo-Wei Zhang$^{1,2,3}$}
\author{Zhen-Qiang Yin$^{1,2,3}$}
\author{Shuang Wang$^{1,2,3}$}
\author{Guang-Can Guo$^{1,2,3}$}
\author{Zheng-Fu Han$^{1,2,3}$}
\affiliation{$^1$CAS Key Laboratory of Quantum Information, University of Science and Technology of China, Hefei 230026, People's Republic of China\\
$^2$Synergetic Innovation Center of Quantum Information $\&$ Quantum Physics, University of Science and Technology of China, Hefei, Anhui 230026, People's Republic of China\\
$^3$State Key Laboratory of Cryptology, P. O. Box 5159, Beijing 100878, People's Republic of China}


\begin{abstract}

The degree of freedom of orbital angular momentum (OAM) is an important resource in high-dimensional quantum information processing, as the quantum number of OAM can be infinite. The Dove prism (DP) is a most common tool to manipulate the OAM light, such as in interferometers. However, the Dove prism does not preserve the polarization of the photon states and decreases the sorting fidelity of the interferometer. In this work, we analyze the polarization-dependent effect of the DP on single-path Sagnac interferometers. The results are instructive to quantum information processing with OAM light. We also proposed a modified single-path beam splitter Sagnac interferometer (BSSI), of which the sorting fidelity is independent on input polarization and can be 100\% in principle. The single-path BSSI is stable for free running. These merits are crucial in quantum information processing, such as quantum cryptography.

\end{abstract}

%\pacs{42.50.Tx, 42.50.Dv, 07.60.Ly}


\maketitle

\section{Introduction}
\label{sect1}

The orbital angular momentum (OAM) of light \cite{Allen1992} is an important resource in many fields, such as quantum simulation \cite{Cardano2015}, quantum metrology \cite{Giovannetti2011,Lavery2013,D’Ambrosio2013,Krenn2016}, micromechanic \cite{Galajda2001,Padgett2011} and high-dimensional quantum information processing \cite{Bourennane2001,Molina-Terriza2007,Barreiro2008,Djordjevic2013,Simon2014,Mirhosseini2015,Sit2016,Nape2016,Erhard2016,Ren2016a,Malik2016,D'Ambrosio2016}. Recently, OAM photon states with quantum number larger than $10^4$ have been created \cite{Fickler2016}. Methods to separate tens of OAM modes simultaneously have also been successfully demonstrated \cite{Berkhout2010,O'Sullivan2012,Mirhosseini2013,Mirhosseini2014} and many other innovative creating and sorting methods have also been proposed \cite{Cai2012,Pu2015,Zhou2016,Walsh2016,Ren2016b}. However, optical interferometers with Dove prisms (DPs) is the dominant method to accomplish the nondestructive sorting of OAM modes \cite{Leach2002,Leach2004,Zhang2014,Malik2016,Erhard2016}. The DP converses the $l$-order OAM state $|l\rangle$ into $exp(i2l\alpha)|-l\rangle$, where $\alpha$ is the rotation angle of the DP. In a Mach-Zehnder interferometer, the OAM mode-dependent phase term $exp(i2l\alpha)$ can be used to separate the $l$-order OAM mode from the $l'$-order one. The original double-path Mach-Zehnder interferometer (MZI) is lack of stability. This shortcoming can be overcome with double and single-path Sagnac interferometers \cite{Erhard2016,Jones2009,Slussarenko2010,D'Ambrosio2012}. The single-path Sagnac interferometer (SPSI) is most robust and can be cascaded for several levels \cite{Zeng2016}. Unfortunately, the DP does not preserve the polarization of the photon states \cite{Padgett1999,Moreno2003}, which decreases the sorting fidelity of the interferometer.

In this work, we analyze the effect of the DP on the single-path Sagnac interferometers. The results indicate that both the output polarization and sorting fidelity of the single-path beam splitter Sagnac interferometer (BSSI) vary with the rotation angle $\alpha$ of the inserted DP. Interestingly, the results also indicate that the polarization of the photon is flipped at the output port with destructive coherence if the input state is horizontal ($|H\rangle$) or vertical ($|V\rangle$) polarized. The flip is independent on the rotation angle $\alpha$. While the sorting fidelity of the single-path polarization beam splitter Sagnac interferometer (PBSSI) depends much less on $\alpha$ and is nearly invariant. However, the single-path PBSSI leads to several percents loss of intensity. By utilizing a simple passive polarization-compensation structure, we implement a high sorting-fidelity BSSI. It can be used as a low-crosstalk OAM sorter. The single-path BSSI is stable for free running. These merits are crucial in quantum information, especially in quantum cryptography.

\section{DP in Single-path Sagnac interferometers}
\label{sect2}

\begin{figure}[htbp]
\centering
\resizebox{6cm}{3cm}{\includegraphics{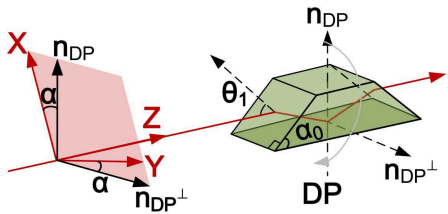}}
\caption{The schematic diagram of a DP. There are two refractive and one total reflective interfaces for a light beam propagating through the DP. $\theta_1$ is the incident angle at the input interface. $\bm{n}_{{}_{DP}}$ is the normal of the base. There is a relative rotation angle $\alpha$ between the lab (with red color) and the DP (with black color) coordinate systems. The rotation of the DP is along Z axis.}
\label{fig1}
\end{figure}

A DP is shaped from a truncated right angle prism with a base angle $\alpha_0$, as shown in Fig. \ref{fig1}. The DP is not polarization preserved and can be expressed by Jones matrices \cite{Padgett1999,Moreno2003} 
\begin{equation}
J_{DP}=R_s'T_{out}R_{inner}T_{in}R_s,
\label{equ1}
\end{equation}
where $T_{in(out)}$ is the refractive matrix of the input (output) surface, $R_{inner}$ is the total internal reflective matrix (Figure \ref{fig1}) and $R_s^{(')}$ is the transformation between the lab and the DP coordinate systems (Figure \ref{fig1}). According to the Fresnel equations, the transmission coefficient of the DP is different for the parallel and perpendicular polarized components and the total internal reflection of the DP will cause a relative phase shift $\Delta\varphi$ between these two components \cite{Born1999}. $\Delta\varphi$ is determined by the incident angle $\theta_3$ and the index of the DP. $J_{DP}$ can then be simplified into the following form
\begin{equation}
\begin{aligned}
J_{DP}(\alpha)=R_s(-\alpha)
\begin{pmatrix}
\sqrt{t_{//}} & 0 \\
0 & \sqrt{t_{\perp}}e^{i\Delta\varphi}
\end{pmatrix}R_s(\alpha),
R_S(\alpha)=
\begin{pmatrix}
cos\alpha & sin\alpha \\
-sin\alpha & cos\alpha
\end{pmatrix}.
\end{aligned}
\label{equ2}
\end{equation}
where $\alpha$ is the angle between the coordinate systems (Figure \ref{fig1}), $t_{//(\perp)}$ is the transmission coefficient of the DP for the polarized light that parallel (perpendicular) to the normal of the base $\bm{n}_{{}_{DP}}$.

\subsection{DP in single-path BSSI}

\begin{figure}[htbp]
\centering
\resizebox{11.69cm}{6.5cm}{\includegraphics{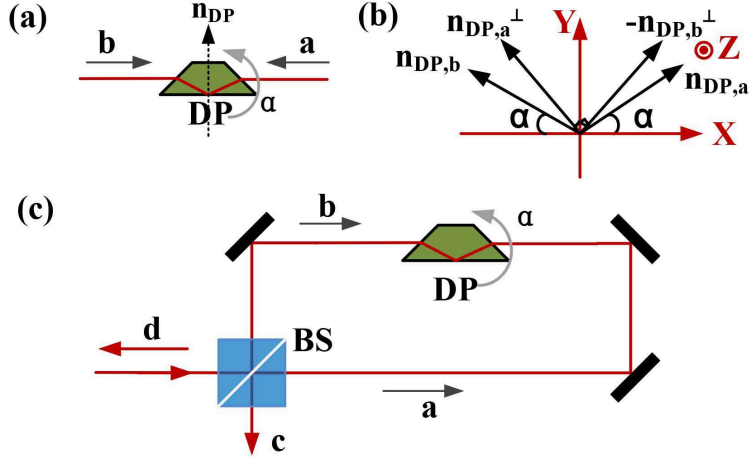}}
\caption{(a) The schematic diagram of a DP with two OAM states incident from opposite directions. (b) The coordinate systems of the DP for light incident from directions a ($n_{{}_{DP,a}}$ and $n^\perp_{{}_{DP,a}}$) and b ($n_{{}_{DP,b}}$ and $n^\perp_{{}_{DP,b}}$). (c) The schematic setup for a single-path BSSI. BS: beam splitter. }
\label{fig2}
\end{figure}

Considering the DP with a rotation angle $\alpha$, two $l$-order OAM states incident from the opposite directions (direction a and b in Figure \ref{fig2}(a)) will experience a relative phase shift $2l(-\alpha)-2l\alpha=-4l\alpha$, as the parallel axes of directions a ($\bm{n}_{{}_{DP,a}}$) and b ($\bm{n}_{{}_{DP,b}}$) are different (see Figure \ref{fig2}(b)). Correspondingly, the coordinate transformations $R_{s,a}$ and $R_{s,b}$ are also different. According to Eq. \ref{equ1}, the polarization variations of these two states become different. Thus, the output state of a single-path BSSI (Figure \ref{fig2}(c)) is expressed as 
\begin{equation}
\begin{aligned}
|\psi\rangle_{out}=&(U^{BS}_p\otimes I_s\otimes I_o)(|b\rangle_p\langle b|\otimes J_{DP}(\pi-\alpha)\otimes P_o(-\alpha))\\
&(|a\rangle_p\langle a|\otimes J_{DP}(\alpha)\otimes P_o(\alpha))(U^{BS}_p\otimes I_s\otimes I_o)|\psi\rangle_{in},
\end{aligned}
\label{equ3}
\end{equation}
where the subscripts $p$, $s$ and $o$ mean the degree of freedom of the path, polarization and OAM, respectively. The input state is expressed as $|\psi\rangle_{in}=|p\rangle_p|s\rangle_s|l\rangle_o$. $I_o$ and $I_s$ are the identity operators. $P_o$ is the phase operator of the DP on OAM states and $P_o(\alpha)|l\rangle_o=exp(i2l\alpha)|l\rangle_o$. $U^{BS}_p$ is the unitary transformation of the BS and the expression is
\begin{equation}
\begin{aligned}
U^{BS}_p=
\begin{pmatrix}
\sqrt{T} & -\sqrt{1-T}\\
\sqrt{1-T} & \sqrt{T}
\end{pmatrix},
\end{aligned}
\label{equ4}
\end{equation}
where $T$ is the transmissivity of the BS.

For simplicity, let $exp(-i4l\alpha)=1$ and $T=1/2$. For a general input state $|\psi\rangle_{in}=(\alpha_0|H\rangle+\beta_0|V\rangle)_s|d\rangle_p|l\rangle_o$, the output state of the single-path BSSI is 
\begin{equation}
\begin{aligned}
|\psi\rangle_{out}=&\frac{e^{2il\alpha}}{2\sqrt{t_{//}+t_{\perp}}}\{(\sqrt{t_{//}}-\sqrt{t_{\perp}}e^{i\Delta\varphi})sin2\alpha(\beta_0|H\rangle_s+\alpha_0|V\rangle_s)|c\rangle_p+[2\alpha_0(\sqrt{t_{//}}cos^2\alpha\\
&+\sqrt{t_{\perp}}e^{i\Delta\varphi}sin^2\alpha)|H\rangle_s+2\beta_0(\sqrt{t_{//}}sin^2\alpha+\sqrt{t_{\perp}}e^{i\Delta\varphi}cos^2\alpha)|V\rangle_s]|d\rangle_p\}|l\rangle_o,
\end{aligned}
\label{equ5}
\end{equation}
where, $|H\rangle$ ($|V\rangle$) represents the polarization that parallel to the X (Y) axis. the complex amplitudes $\alpha_0$ and $\beta_0$ satisfy $\alpha_0^2+\beta_0^2=1$. In the case of $exp(-i4l\alpha)=-1$, the output state can be obtained by exchanging the output ports c and d of Equation \ref{equ5}. According to Equation \ref{equ5}, the output polarizations of both ports are changed. The output polarization with destructive coherence only determined by the input polarization: $\alpha_0|H\rangle+\beta_0|V\rangle\xrightarrow{}\beta_o|H\rangle+\alpha_0|V\rangle$. The output polarization with destructive coherence is complex and depends on the input polarization, the DP and the rotation angle. Here we discuss two special cases: $\alpha_0=1$ or $\beta_0=1$. In these cases, the polarization at the output port with constructive coherence is preserved. While the polarization at the output port with destructive coherence is flipped. The output polarizations become independent on the DP and $\alpha$.

\begin{figure}[htbp]
\centering
\resizebox{8cm}{5.5cm}{\includegraphics{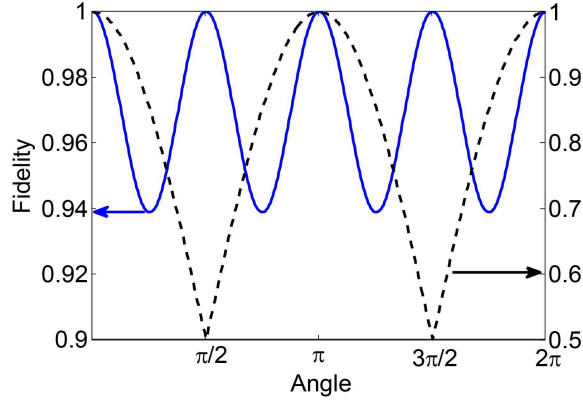}}
\caption{The sorting fidelities of the single-path BSSI with the rotation angle $\alpha$ (the blue line) and the relative phase $\Delta\varphi$ added by the DP (the black dashed line), respectively. The input polarization is $|H\rangle$. For the blue line, the DP-dependent parameters are measured values of a practical DP ($t_{//}=0.9877,t_{\perp}=0.9475,\Delta\varphi=0.159\pi$). For the black dashed line, the rotation angle is set as $\alpha=\pi/4$.}
\label{fig3}
\end{figure}

The sorting fidelity of the single-path BSSI for OAM modes is 
\begin{equation}
\mathcal{F}=P_{constructive},
\label{equ6}
\end{equation}
where $P_{constructive}$ is the output probability of the state at the port with constructive coherence. According to Equation \ref{equ5}, once the fabrication parameters of the DP is determined, $P_c=|{}_p\langle c|\psi\rangle_{out}|^2$ is independent on the input polarization and only varies with the rotation angle $\alpha$. Moreover, when $t_{//}=t_{\perp}$, $P_d$ is also independent on  the input polarization. By taking into account the coating technology, $t_{//}-t_{\perp}$ is at the order of $10^{-2}$. Thus, both $P_c$ and $P_d$ nearly only vary with $\alpha$. The blue line in Figure \ref{fig3} gives the sorting fidelity of the single-path BSSI with rotation angle $\alpha$. The parameters of the DP ($t_{//}=0.9877,t_{\perp}=0.9475$ and $\Delta\varphi\simeq 0.159\pi$) are measured values. The minimum $\mathcal{F}$ is about 93.9\%.

According to Equations \ref{equ5} and \ref{equ6}, the dip of the sorting fidelity depends on the relative phase shift $\Delta\varphi$ of the DP. The black dashed line in Figure \ref{fig3} gives the minimum sorting fidelity ($\alpha=\pi/4$) with varying $\Delta\varphi$, where the input polarization is set as $|H\rangle$. The sorting fidelity is 1 and 0.5 when $\Delta\varphi$ is $(2n+1)\pi/2$ and $(2n+1)\pi/2$, respectively. What should be noted is that $\Delta\varphi$ is determined by the incident angle and the refractive index of the DP. Although $\Delta\varphi\in [0,2\pi]$ in Figure \ref{fig3}, a $\pi/2$ relative phase shift requires the refractive index of the DP being at least 2.41. Thus, for common cases, $\Delta\varphi$ is less than $\pi/2$. For example, if the refractive index is 1.52 (BK7 glass), the maximum $\Delta\varphi$ is $0.259\pi$.

\subsection{DP in single-path PBSSI}

The single-path PBSSI (by replacing the BS with a PBS in Figure \ref{fig2}(c)) is different from the BSSI. If the DP is moved away, the state will output from port c only. Although the DP changes the polarization, the polarization beam splitter (PBS) then acts as a filter so that the polarization output from direction a (b) is pure $|H\rangle$ ($|V\rangle$). The PBSSI usually works for polarization states in the form of $\frac{1}{\sqrt{2}}(|H\rangle+e^{i\varphi}|V\rangle)$. When the OAM mode-dependent phase shift satisfies $e^{-i4l\alpha}\cdot e^{i\varphi}=\pm 1$, the single-path PBSSI works as an OAM sorter. According to Equation \ref{equ2}, for the single-path PBSSI, the DP works as follows:
\begin{equation}
\begin{aligned}
&|H\rangle\xrightarrow{DP,a}(\sqrt{t_{//}}cos^2\alpha+\sqrt{t_{\perp}}sin^2\alpha e^{i\Delta\varphi})|H\rangle+(\sqrt{t_{//}}-\sqrt{t_{\perp}}e^{i\Delta\varphi})sin\alpha cos\alpha|V\rangle,\\
&|V\rangle\xrightarrow{DP,b}(\sqrt{t_{//}}sin^2\alpha+\sqrt{t_{\perp}}cos^2\alpha e^{i\Delta\varphi})|V\rangle-(\sqrt{t_{//}}-\sqrt{t_{\perp}}e^{i\Delta\varphi})sin\alpha cos\alpha|H\rangle.
\end{aligned}
\label{equ7}
\end{equation}
Then, the output state becomes 
\begin{equation}
\begin{aligned}
|\psi\rangle_{out}=&\frac{e^{2il\alpha}}{\sqrt{2(t_{//}+t_{\perp})}}\{[(\sqrt{t_{//}}cos^2\alpha+\sqrt{t_{\perp}}sin^2\alpha e^{i\Delta\varphi})|H\rangle\pm(\sqrt{t_{//}}sin^2\alpha\\
&+\sqrt{t_{\perp}}cos^2\alpha e^{i\Delta\varphi})|V\rangle]|c\rangle_p+(\sqrt{t_{//}}-\sqrt{t_{\perp}}e^{i\Delta\varphi})sin\alpha cos\alpha(|V\rangle\mp|H\rangle)|d\rangle_p\}|l\rangle_o.
\end{aligned}
\label{equ8}
\end{equation}
According to Equation \ref{equ8}, the state is partially output from port d and decreases the output intensity at port c. The state output from port c is OAM mode-dependent. For two OAM modes satisfying $4(l_1-l_2)\alpha=\pi$, the overlap of the output polarizations at port c is
\begin{equation}
\begin{aligned}
P_{overlap}=&|\langle \psi(l_1)_{out,c}|\psi(l_2)_{out,c}\rangle|^2\\
=&\frac{1}{N^2}\cdot|[(\sqrt{t_{//}}cos^2\alpha+\sqrt{t_{\perp}}sin^2\alpha e^{-i\Delta\varphi})\langle H|-(\sqrt{t_{//}}sin^2\alpha+\sqrt{t_{\perp}}cos^2\alpha e^{-i\Delta\varphi})\langle V|]\\
&[(\sqrt{t_{//}}cos^2\alpha+\sqrt{t_{\perp}}sin^2\alpha e^{i\Delta\varphi})|H\rangle+(\sqrt{t_{//}}sin^2\alpha+\sqrt{t_{\perp}}cos^2\alpha e^{i\Delta\varphi})|V\rangle]|^2\\
=&\frac{1}{N^2}(t_{//}-t_{\perp})^2cos^2(2\alpha),
\end{aligned}
\label{equ9}
\end{equation}
where $N=(t_{//}+t_{\perp})(1-\frac{1}{2}sin^2(2\alpha))+\sqrt{t_{//}t_{\perp}}sin^2(2\alpha)cos\Delta\varphi$ is the renormalized factor at port c. 

\begin{figure}[htbp]
\centering
\resizebox{7.82cm}{5.5cm}{\includegraphics{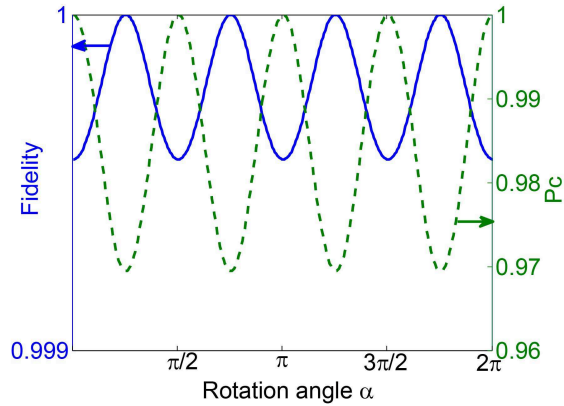}}
\caption{The sorting fidelity (the blue line) and the output probability $P_c$ of the state at port c (the green dashed line) of the single-path PBSSI with the rotation angle $\alpha$. The input polarization is $|H\rangle$. The DP-dependent parameters are $t_{//}=0.9877,t_{\perp}=0.9475,\Delta\varphi=0.159\pi$.}
\label{fig4}
\end{figure}

Hence, the sorting fidelity of the single-path PBSSI is 
\begin{equation}
\mathcal{F'}=1-P_{overlap}.
\label{equ10}
\end{equation}
Different from that in the single-path BSSI, the sorting fidelity is 100\% when $\alpha=(2n+1)\pi/4$ while becomes minimum when $\alpha=2n\pi/4$. As $t_{//}-t_{\perp}$ is at the order of $10^{-2}$, the polarization overlap of a PBS Sagnac interferometer (Equation \ref{equ9}) is negligible. As shown in Figure \ref{fig4}, the sorting fidelity (the green dashed line) of a practical DP is larger than 0.9995 for arbitrary $\alpha$. Although the polarization at port c is elliptically polarized, the combination of one HWP and two QWPs is enough to rotate the light into linear polarization. Thus, the sorting fidelity of the single-path PBSSI is unaffected by the DP. The only concern is the intensity loss at port d. The green dashed line in Figure \ref{fig4} gives the output probability of the photon at port c. The total output probability of ports c and d is renormalized as 1. The maximum intensity loss is about 3\%. It should be noted that the sorting fidelity and intensity loss also depends on parameters of the DP. However, for a commercial DP, the loss is usually within several percents and is acceptable.

\section{High sorting-fidelity single-path BSSI for OAM states}
\label{sect3}

As discussed in Section \ref{sect2}, the sorting fidelity of the single-path PBSSI for OAM modes is almost unaffected by the DP. Although possessing high sorting fidelity for OAM modes, the PBSSI is only practicable for polarization with the certain form $\frac{1}{\sqrt{2}}(|H\rangle+exp(i\varphi)|V\rangle)$. While the minimum sorting fidelity of a single-path BSSI is about 93.9\%, which means the single-path BSSI alone leads to about 6\% sorting error for OAM modes. This is unaccepted in some applications, such as in quantum key distribution (QKD) with BB84 protocol. The total system error in the BB84 protocol should be less than 11\% \cite{Shor2000}. 
In this section, we first measured the parameters of a practical DP so that we can estimate the sorting fidelity of the original single-path BSSI in Figure \ref{fig2}(c) for OAM modes. Then, we proposed a modified single-path BSSI to get high sorting fidelity for OAM modes.

\begin{figure}[htbp]
\centering
\resizebox{12cm}{2.95cm}{\includegraphics{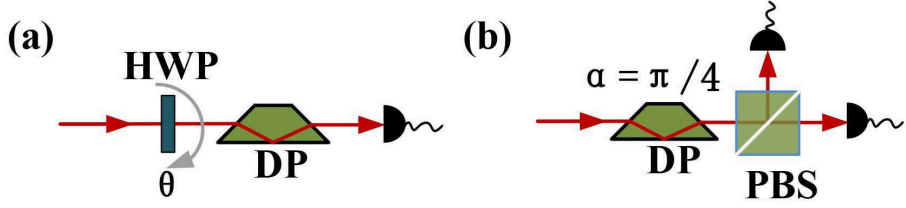}}
\caption{The device for measuring (a) the transmissivity coefficients $\sqrt{t_{//}}$, $\sqrt{t_{\perp}}$ and (b) the relative phase shift $\Delta\varphi$ of the DP. PBS: polarizing beamsplitter.}
\label{fig5}
\end{figure}

A measurement device showing in Figure \ref{fig5}(a) is implemented to determine $J_{DP}$ of a practical DP (the base angle is $45^{\circ}$ and the length is 63 $mm$). We rotate the polarization of the light to fulfill the measurement. The method is equivalent to rotate the DP but easier to implement and more precise. By rotating a HWP before the DP, the output light intensity varies and the parameters $t_{//}$ and $t_{\perp}$ can be determined. The measurement gives that $t_{//}\simeq 0.987$ and $t_{\perp}\simeq 0.945$.

Figure \ref{fig5}(b) gives the experimental setup to measure to relative phase shift $\Delta\varphi$. The rotation angle $\alpha$ of the DP is $\pi/4$. For an input state $|H\rangle$, the output state becomes
\begin{equation}
\begin{aligned}
|\psi\rangle_{out}=\sqrt{\frac{1}{N}}J_{DP}(\pi/4)|H\rangle=\frac{1}{2\sqrt{N}}[(\sqrt{t_{//}}+\sqrt{t_{\perp}}e^{i\Delta\varphi})|H\rangle+(\sqrt{t_{//}}-\sqrt{t_{\perp}}e^{i\Delta\varphi})|V\rangle],
\end{aligned}
\label{equ13}
\end{equation}
where the normalization factor $N=(t_{//}+t_{\perp})/2$. The output polarization is then analyzed by a PBS. Thus, $\Delta\varphi$ can be calculated by measuring the output intensities after the PBS. The measurement result is $P_{out,H}^e\simeq 0.939$. Then we obtain $\Delta\varphi\simeq 0.159\pi$. This completes the measurement of $J_{DP}$. According to Equations \ref{equ5} and \ref{equ6}, the sorting fidelity of the original single-path BSSI with the practical DP here can not be higher than 93.9\% when $\alpha=\pi/4$.

\begin{figure}[htbp]
\centering
\resizebox{12cm}{6.8cm}{\includegraphics{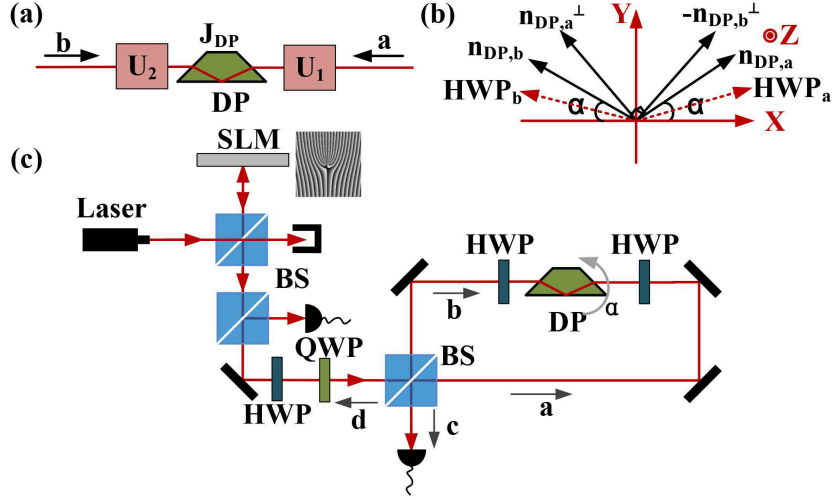}}
\caption{(a) The unitary transformations to rotate the polarization. (b) $n_{DP,a(b)}$ and $n^{\perp}_{DP,a(b)}$ consist the coordinate system of the DP for direction a (b). $HWP^{a(b)}_i$ is the fast axis of the half-wave plate in the BSSI for direction a (b). (c) The proof-in-principle experimental setup of the modified single-path BSSI. SLM: spatial light modulator; QWP: quarter-wave plate.}
\label{fig6}
\end{figure}

According to Equation \ref{equ3}, the sorting fidelity decreases because of the asymmetry of coordinate transformation $R_s$ for the opposite directions in the Sagnac loop. If we rotate the lab coordinate system to be coincident with the DP coordinate system, then there is no difference for polarization in both directions. An equivalent way is to rotate the polarization $\alpha_0|H\rangle+\beta_0|V\rangle$ into the form $\pm\alpha_0|//\rangle_{a(b)}+\beta_0|\perp\rangle_{a(b)}$ before incident into the DP, where $|//\rangle_{a(b)}$ ($|\perp\rangle_{a(b)}$) is the polarization that is parallel (perpendicular) to $\textbf{n}_{{}_{DP,a(b)}}$. That is, as shown in Figure \ref{fig6}(a), if the transformations in the degree of freedom of polarization satisfy
\begin{equation}
\begin{aligned}
U_{1,a}(\alpha_0|H\rangle+\beta_0|V\rangle)&=\pm\alpha_0|//\rangle_{a}+\beta_0|\perp\rangle_{a},\\
U_{2,b}(\alpha_0|H\rangle+\beta_0|V\rangle)&=\pm\alpha_0|//\rangle_{b}+\beta_0|\perp\rangle_{b},\\
U_{2,a}J_{DP}(\alpha)U_{1,a}&=U_{1,b}J_{DP}(\pi-\alpha)U_{2,b},
\end{aligned}
\label{equ11}
\end{equation}
then the polarizations output from both directions are the same and the sorting fidelity is independent on polarization, where the subscript $a$ ($b$) denotes the rotation transformation in direction a (b), as the rotation transformation $U_{1(2)}$ in direction a is usually not equivalent to that in direction b. The solution for Equation \ref{equ11} is not unique, a simple solution is 
\begin{equation}
\begin{aligned}
U_{1,a}&=U_{2,a},U_{1,b}=U_{2,b},U_{2,b}=U_{2,b}(\alpha)=U_{1,a}(-\alpha).\\
U_{1,a}&=U_{1,a}(\alpha)=
\begin{pmatrix}
cos(\alpha) & sin(\alpha)\\
sin(\alpha) & -cos(\alpha)
\end{pmatrix},\\
\end{aligned}
\label{equ12}
\end{equation}
The unitary transformation of Equation \ref{equ12} can be easily realized by a HWP. $U_{1(2),a}(\alpha)$ represents a HWP with its fast axis rotating the angle $\alpha/2$ at its horizontal axis ($HWP_a$ in Figure \ref{fig6}(b)). In the opposite direction, the rotation angle of the fast axis of the same HWP becomes $\pi-\alpha/2$ ($HWP_b$ in Figure \ref{fig6}(b)). This corresponds to $U_{1(2),b}$. Thus, only two more HWPs are necessary to complete the modified single-path BSSI, as shown in Figure \ref{fig6}(c). Thus, for a general input state $\alpha_0|H\rangle+\beta_0|V\rangle$, the output state in both directions a and b is $\frac{1}{\sqrt{t_{//}\alpha^2_0+t_{\perp}\beta^2_0}}(\sqrt{t_{//}}\alpha_0|H\rangle-\sqrt{t_{\perp}}\beta^2_0e^{i\Delta\varphi}|V\rangle$. Hence, the sorting fidelity of the single-path BSSI is independent on the input polarization and can be 100\% in principle, though the polarization is not preserved. However, parameters $t_{//},t_{\perp}$ and $\Delta\varphi$ are fixed. Hence, passive optical elements after the single-path BSSI are enough to compensate the polarization variation, if necessary.

The proof-in-principle experimental setup of the single-path BSSI for OAM photon states is demonstrated in Figure \ref{fig6}(c). As the first-order coherence of a single photon can be simulated by coherence light, a laser diode is used as the light source. The corresponding wavelength is 780 $nm$. The OAM modes are generated by the spatial light modulator (SLM). The OAM modes are then input into the single-path BSSI. The transmissivity of the BS is 0.52. The rotation angle of the DP is $\pi/4$. As the path of output port d overlaps the input path, the corresponding intensity is detected by inserting a BS into the optical path. The intensities of both ports are detected by power meters. The HWP and quarter-wave plate (QWP) before the BSSI is used to verify the polarization-independent property of the single-path BSSI. As a proof-in-principle experiment, only six polarizations are verified: $|H\rangle$, $|V\rangle$, $|+\rangle=|H\rangle+|V\rangle$, $|-\rangle=|H\rangle-|V\rangle$, $|L\rangle=|H\rangle-i|V\rangle$ and $|R\rangle=|H\rangle+i|V\rangle$.

\begin{table}
\centering
\caption{The sorting fidelities of the modified single-path BSSI for different polarizations.}
\begin{tabular*}{1\textwidth}{@{\extracolsep{\fill}}cccccccccccc}
%\begin{tabular}{cccccccccccc}
\hline \hline

\multicolumn{2}{c}{\bf $l=$} & 1 & 2 & 3 & 4 & 5 & 6 & 7 & 8 & 9 & 10 \\
\hline 
\multirow{6}*{$\mathcal{F}$} & $|H\rangle$ & 0.982 & 0.978 & 0.980 & 0.974 & 0.976 & 0.969 & 0.967 & 0.963 & 0.957 & 0.956 \\
 & $|V\rangle$ & 0.981 & 0.974 & 0.980 & 0.971 & 0.975 & 0.962 & 0.965 & 0.954 & 0.955 & 0.944 \\ 
 & $|+\rangle$ & 0.981 & 0.982 & 0.982 & 0.980 & 0.977 & 0.975 & 0.971 & 0.970 & 0.961 & 0.963 \\
 & $|-\rangle$ & 0.978 & 0.973 & 0.979 & 0.970 & 0.974 & 0.964 & 0.965 & 0.956 & 0.954 & 0.950 \\
 & $|L\rangle$ & 0.976 & 0.977 & 0.977 & 0.974 & 0.970 & 0.967 & 0.960 & 0.959 & 0.947 & 0.951 \\
 & $|R\rangle$ & 0.980 & 0.983 & 0.979 & 0.978 & 0.972 & 0.970 & 0.961 & 0.962 & 0.948 & 0.954 \\

\hline \hline
\end{tabular*}
%\arrayrulewidth=2pt
\label{tab:1}
\end{table}

Table \ref{tab:1} gives the sorting fidelities of the modified single-path BSSI for ten OAM modes and six polarizations. All sorting fidelities in Table \ref{tab:1} are larger than 0.939 and the highest sorting fidelity is about 0.982. There are no significant differences of the sorting fidelities for different polarizations. These results prove that the modified single-path BSSI is independent on the input polarization and improves the sorting fidelity significantly. The sorting fidelity decreases as the OAM mode $l$ increases. This is due to the limited rotation precision of the rotators used in the experiment. The rotation precision is only 2 degree. The phase error of the mode-dependent term of the BSSI increases with $l$. And the corresponding sorting fidelity decreases. Another reason that limits the sorting fidelity is phase only SLM. In order to generate high-purity OAM mode, both phase and amplitude of the light should be modulated. Only the phase is modulated in the experiment (Figure \ref{fig6}(c)). The generation purity decreases as the OAM mode increases. This is why the highest sorting fidelity is about 0.982. The sorting fidelity can be improved further by using rotators with higher rotation precision. It can also be improved by improving the phase pattern added on the SLM \cite{Clark2016}. As the single-path Sagnac interferometer is stable for free running, the standard deviations of the sorting fidelities in Table \ref{tab:1} are all less than 0.2\% without any active feedback control. Thus, the proposed structure is promising for practical applications of OAM photon states.

\section{Conclusion}

In conclusion, we have given a detailed analysis of polarization-dependent properties of the single-path BSSI and PBSSI for OAM photon states. The analysis is instructive to quantum information processing with OAM photon states. We have also demonstrated a high sorting-fidelity single-path BSSI that is independent on input polarizations. The modified single-path BSSI is stable for free running. Thus, the proposed structure is promising for practical high-dimensional quantum information processing.

\section*{Acknowledgments}
This work has been supported by the National Natural Science Foundation of China (Grant Nos. 61675189, 61627820, 61622506, 61475148, 61575183), the National Key Research And Development Program of China (Grant Nos.2016YFA0302600, 2016YFA0301702), the "Strategic Priority Research Program(B)" of the Chinese Academy of Sciences (Grant No. XDB01030100).

\end{document}